\documentclass[letterpaper, 10 pt, conference]{ieeeconf}

\IEEEoverridecommandlockouts

\usepackage{color}
\usepackage{enumerate,graphicx,epstopdf}
\usepackage{amsmath,amssymb,bm,amsthm}
\usepackage{cite}
\usepackage{mathtools}
\usepackage{array}
\usepackage{algpseudocode,algorithm,algorithmicx}
\usepackage{color}
\usepackage{booktabs}
\usepackage[hyphens]{url}
\usepackage{lipsum}
\usepackage[dvipsnames]{xcolor}
\usepackage[deletedmarkup=sout,authormarkup=superscript]{changes}
\usepackage[normalem]{ulem}

\newtheorem{rmk}{Remark}

\newcommand{\reals}{\mathbb{R}}

\newcommand{\mc}[1]{\mathcal{#1}}

\newcommand{\diag}[0]{\mathrm{diag}~}

\definechangesauthor[name={Efe}, color=NavyBlue]{EB}
\definechangesauthor[name={Dominic}, color=RedViolet]{DLM}
\definechangesauthor[name={Alisa}, color=OliveGreen]{AR}

\begin{document}
\title{In-layer Thermal Control of a Multi-layer Selective Laser Melting Process
\thanks{The authors are with the ETH Z\"{u}rich Automatic Control Laboratory, Physikstrasse 3, 8092 Z\"{u}rich, Switzerland. Emails: \texttt{\{dliaomc, ebalta, wueestry, ralisa, jlygeros\}@ethz.ch}. A. Rupenyan is also with Inspire AG. Research supported by NCCR Automation, a National Centre of
 		Competence in Research, funded by the Swiss National Science
 		Foundation (grant number 180545).
%  		This research is supported by the Swiss National Science Foundation through NCCR Automation. 
}}
\author{Dominic Liao-McPherson, Efe C. Balta, Ryan W\"{u}est, Alisa Rupenyan, and John Lygeros}

\maketitle

\begin{abstract}
Selective Laser Melting (SLM) is an additive manufacturing technology that builds three dimensional parts by melting layers of metal powder together with a laser that traces out a desired geometry. SLM is popular in industry, however the inherent melting and re-solidification of the metal during the process can, if left uncontrolled, cause excessive residual stress, porosity, and other defects in the final printed parts. This paper presents a control-oriented thermal model of a multi-layer SLM process and proposes a structured model reduction methodology with an associated reduced order model based in-layer controller to track temperature references. Simulation studies demonstrate that the controller is able to prevent layer-to-layer heat buildup and that good closed-loop performance is possible using relatively low-order models.
\end{abstract}

%%*************************************

\section{Introduction}

Additive manufacturing (AM) provides flexibility and customizability to produce intricate product geometries with a wide variety of materials.
While AM processes are widely adopted in practice, reliability and repeatability are often important challenges due to the complex physics involved in the processes and disturbances that affect process quality~\cite{dowling2020review}.
In this work, we focus on Selective Laser Melting (SLM), and propose a multi-layer control oriented model together with an in-layer state feedback controller for temperature regulation. 
SLM is a popular AM process that is widely adopted in industry~\cite{yap2015review}. In SLM, metallic powder is deposited in a thin layer, before a laser moves along a predetermined path to melt portions of the powder layer and fuse it with the underlying layer or substrate. 
This process is repeated over several layers to gradually build up a 3D part.
Thermal stresses due to melting and solidification may induce excessive residual stresses on the resulting part \cite{li2018modeling}. 
Additionally, excessive heat accumulation in subsequent layers due to the melting process may cause defects such as porosity, or disturbances in the melt pool \cite{kasperovich2016correlation,shi2016effects}, which in turn affect the mechanical properties of the resulting 3D part.
Therefore, closed-loop process controllers have the potential to significantly improve performance and printed part quality.

An important challenge when developing closed-loop control for the SLM process is a lack of control-oriented models~\cite{al2021control}.
An effective way of controlling heat accumulation is by adjusting laser power to influence the process temperature\cite{al2021control}.
Data-driven concepts using multiple in-situ sensors for closed-loop control are presented in \cite{renken2017development}, but require large training data sets.
Accurate analytical and computational models to simulate the process dynamics are proposed in~\cite{ning2019analytical,yang2020computationally,irwin2021iterative}. However, these models are typically not suitable for control design as they are often computationally expensive.
Additionally, none of the previously mentioned models capture the multi-layer dynamics of the process, which is a crucial aspect for accurate process control.
A control-oriented process model considering loose powder or bulk material in a preceding layer is developed in~\cite{shkoruta2020experimental}, but the thermal distribution of the preceding layer and heat accumulation is not considered. 

Recent work has proposed methods to perform closed-loop control for SLM processes.
In~\cite{spector2018passivity}, a graph-based model for the thermodynamics within a single layer is used for iterative learning control. 
Despite promising results for the given case, the model cannot capture heat accumulation as it only considers a single layer over a substrate with uniform temperature distribution.
In \cite{shkoruta2021real}, a proportional-integral (PI) controller is considered for in-layer control of melt pool size through laser power.
Similarly, simulated pyrometer measurements are used in a proof of concept proportional controller in~\cite{renken2018model}. 
While demonstrating practical applicability, these controllers do not consider optimal regulation or input constraints of the physical process, and do not capture the multi-layer process dynamics.
In~\cite{asadi2021gaussian}, a Gaussian Process surrogate model for a single layer process is considered for in-layer closed loop model predictive control (MPC) of the laser power. 
While MPC provides improved control performance, multi-layer dynamics of the process and heat accumulation from previous layers again is not considered.  
A layer-to-layer control oriented-model for SLM is presented in~\cite{wang2020layer}.
The model keeps track of a fixed number of top layers. 
However, the performance and approximation error of this model reduction is not demonstrated and no controller is presented with the proposed model.

This work presents a control-oriented multi-layer model that captures the layer-to-layer thermodynamics of SLM. 
We utilize a graph-based approach to model the thermal dynamics of the printed part over multiple layers, building on the previous work in \cite{spector2018passivity} for a single layer.
We then propose a structured methodology to construct a reduced-order model (ROM) and quantify the approximation error of the proposed ROM through a representative case study.
Using the multi-layer ROM, we develop an in-layer finite-horizon reduced-order linear quadratic regulator (LQR) that uses state feedback to track desired temperature references.
To the best of the authors' knowledge, this work is the first effort to model the multi-layer thermodynamics of the SLM process in the context of closed-loop in-layer process controllers.
The contributions of this work are threefold:
\begin{enumerate}[(i)]
    \item A multi-layer control-oriented model for the layer-to-layer thermal dynamics of an SLM process.
    \item Reduced order approximations of the multi-layer model, considering the thermal properties of the printed layers in the process.
    \item A preliminary state feedback controller for in-layer control to achieve uniform temperature gradients in subsequent layers.
\end{enumerate}

The paper is organized as follows. In Section~\ref{ss:problem-formulation} we formalize the multi-layer thermal control problem and derive full and reduced-order models in Section~\ref{ss:modelling}. Section~\ref{ss:feedback-control} presents an in-layer controller based on the reduced order model, Section~\ref{ss:simulations} illustrates the performance of the controller through simulated case studies, and Section~\ref{ss:conclusion} concludes the paper and discusses directions for future work.

\section{Problem Formulation} \label{ss:problem-formulation}
We consider an SLM process with $N$ layers indexed by $k$. While the physics of SLM processes are complex and include a variety of phenomena\footnote{E.g., phase transitions, meltpool convection, atmospheric convection, plasma generation, powder gain kinetics and so on.}, successful models of the temperature field omitting many of these phenomena have been reported~\cite{spector2018passivity,gusarov2007heat,verhaeghe2009pragmatic}. Adopting similar assumptions, we consider the volumetric enthalpy $H_k$ and temperature $T_k$ of the process. These fields evolve in a rectangular build volume $\Omega_k \subset \reals^{3}$ and are governed by the heat equation
\begin{equation} \label{eq:pde-heat-equation}
    \frac{d}{dt}H_k(x,t)\! =\! \vec{\nabla} \cdot (\kappa(x) \nabla T_k(x,t)),~x \in \Omega_k,~t\in [0,\tau_k],
\end{equation}
where $\kappa$ is the thermal conductivity and $\tau_k$ is the layer print time. Following \cite{spector2018passivity}, we neglect phase transition effects and let $\textstyle{H = c_p T}$, where $c_p$ is the specific heat capacity of the material. The temperature field is subject to boundary conditions $\mc{P}(\nabla T, T,q) = 0$ on $\partial\Omega_k$. These boundary conditions include a fixed temperature on the lower surface of the volume (i.e., the build plate is held at a fixed temperature $T_s$), convective cooling on the upper surface, and heat flux $q$ applied by the laser.

After each layer is printed, the build plate is lowered to accommodate a new layer, enlarging the build volume, and a fresh layer of power is spread via the \emph{recoating} process. During this period, the build volume cools and heat flows into the new powder layer. These steps are represented by the relation
\begin{equation}\label{eq:layerOp}
    T_{k+1}(x,0) = \mc{S}_k(T_{k}(x,\tau_k),\tau_c),
\end{equation}
where $\tau_c$ is the recoating time.

The laser moves along a trajectory $\textstyle{\mu_k:[0,\tau_k]\to\reals^2}$ that defines the sliced geometry of the part at layer $k$. As in \cite{spector2018passivity,asadi2021gaussian} we adopt a Gaussian beam model and take the laser power as our controlled variable. Specifically, the heat flux due to the laser is
\begin{equation}
    q(x,t) = \mc{B}(x,t) u(t)
\end{equation}
where $x\in \reals^2$ is the two-dimensional spatial variable,
\begin{equation} \label{eq:gaussian_power_dist}
    \mc{B}(x,t) = \frac{\alpha}{\sqrt{|\Sigma|(2\pi)^2}} e^{-\frac12\|x-\mu(t)\|_{\Sigma^{-1}}^2}
\end{equation}
is the beam intensity, $\alpha \in (0,1)$ is the flux fraction absorbed by the material, and $u$ is the laser power. The covariance matrix $\Sigma = \frac{R^2}{9} I$ is chosen so as $99.7\%$ of the laser power is deposited within the radius $R$ of the beam.

In this paper, we choose the output
\begin{equation}
    y(t) = \int_{\partial\Omega^s_k} \mc{B}(x,t)^T T(x,t)~dx,
\end{equation}
which is a weighted average over the top surface $\partial \Omega_k^s$ of the temperature under the laser.  Our objective is to construct a feedback controller that drives $y$ to a reference trajectory $y^d_k$ during each layer.

\begin{figure}[htbp]
	\centering
	\includegraphics[width=0.95\columnwidth]{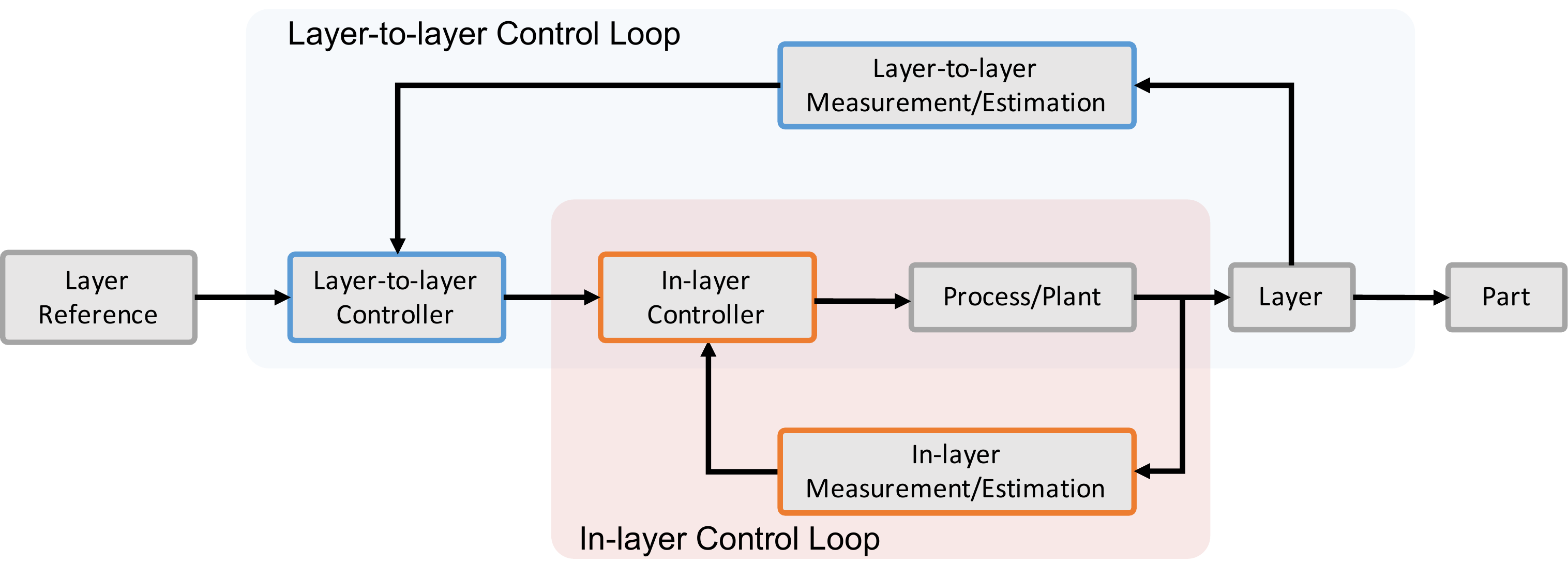}
	\caption{A typical two-loop architecture for SLM control. The inner-loop operates at high ($1$ to $10$ kHz) rates while the layer-to-layer (L2L) loop executes once per layer, typically in the $0.01$ to $10$ Hz range.}
	\label{fig:block-diagram}
\end{figure}

A typical SLM control architecture consists of two feedback loops as illustrated in Figure~\ref{fig:block-diagram}. The inner in-layer loop adjusts the power at run-time, i.e., while the layer is printed, while the outer layer-to-layer (L2L) loop provides input profiles for a complete layer. In this paper, we focus on in-layer control for improving the closed-loop performance through efficient state feedback updates.

\section{Multi-layer Thermal Modelling} \label{ss:modelling}
The multi-layer control problem involves a partial differential equation (PDE), which is spatially discretized for computational tractability. In this section, we extend the single-layer graph-based modelling/discretization framework from \cite{spector2018passivity} to consider multiple layers and use it to derive a simulation and a reduced-order model for control design.

\begin{figure}[htbp]
    \centering
    \includegraphics[width=0.6\columnwidth]{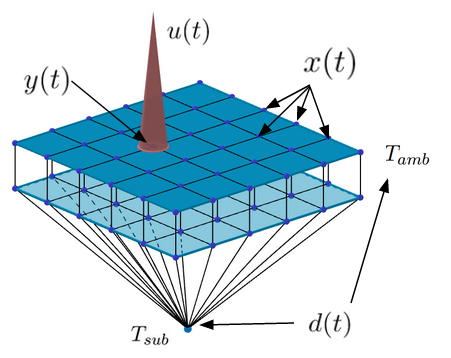}
    \caption{The build volume is descritized using a grid. The system state is the temperature of all the cells, the output is a weighted average of the temperature below the laser, the control input is the laser power, and ambient and substrate temperature act as exogenous inputs.}
    \label{fig:grid_diagram}
\end{figure}

\subsection{Multi-layer Simulation Model}
To discretize the build volume, we construct a grid as shown in Figure~\ref{fig:grid_diagram}. As in \cite{spector2018passivity}, each node represents a rectangular cell of size $\delta x \times \delta y \times \delta z$. These nodes are used to define a weighted graph $\mc{G} = (\mc{V},\mc{E},\kappa)$ where $\mc{V}$ are the nodes, $\mc{E}$ are the edges, and $\kappa^{ij}$ is the conductivity\footnote{The difference in conductivity of powder and solidified metal is accounted for using the method in \cite{roy2018heat}.} between nodes $i$ and $j$. The equation for each node in the first layer is 
\begin{multline}\label{eq:single_layer}
    C_{p}^i \dot{T}^i(t) = -\sum_{j\in \mc{N}_i} \kappa^{ij} (T^i(t) - T^j(t)) + \kappa_s(T_s - T_i(t))\\ + h_{\infty}(T_\infty - T^i(t)) + b^i(t)u(t)
\end{multline}
where $\mc{N}_i$ are the neighbours of node $i$, $C_p^i$ is the heat capacity of node $i$, $b^i(t)$ is the Gaussian power distribution \eqref{eq:gaussian_power_dist} evaluated at node $i$, $\kappa_s$ is the thermal conductivity between nodes and the build plate, $T_s$ is the temperature of the build plate, $h_\infty$ is the convection coefficient between nodes and the atmosphere, and $T_\infty$ is the temperature of the ambient atmosphere.

To extend the model \eqref{eq:single_layer} into a multi-layer model, the graph $\mc{G}_k$ is made layer dependent and is augmented with $\ell^i_k$, the layer number of node $i$, i.e., $\mc{G}_k = (\mc{V}_k,\mc{E}_k,\kappa_k,\ell_k)$. The temperature evolution equation for node $i$ is then
\begin{subequations} \label{eq:EOM_graph}
\begin{multline}\label{eq:multi-1}
C_{p}^i \dot{T}^i(t) = -\sum_{j\in \mc{N}_i} \kappa^{ij} (T^i(t) - T^j(t)) + \delta_k^i \kappa_s(T_s - T_i(t))\\ + \sigma_k^i h_{\infty}(T_\infty - T^i(t)) + \delta_k^i b^i_k(t)u(t)
\end{multline}
where the binary variables
\begin{equation}\label{eq:multi-2}
    \sigma_k^i = \begin{cases}
    1 & \text{ if } \ell^i_k = k\\
    0 & \text{ else }
    \end{cases} \text{   and   } \delta_k^i = \begin{cases}
    1 & \text{ if } \ell^i_k = 1\\
    0 & \text{ else }
    \end{cases},
\end{equation}
\end{subequations}
track if a given node is in the top most or bottom most layer and $b^i_k(t)$ is the beam intensity function \eqref{eq:gaussian_power_dist} evaluated at the centre of node $i$.

Let $m_k = {|\mc{V}_k|}$ denote the number of nodes, and $x_k = [T^1~~\ldots~~T^{m_k}]^T$  the state of the system, i.e., the collected temperature of all nodes. Further, introduce the binary vectors $\sigma_k = [\sigma_k^1~~\ldots~~\sigma_k^{m_k}]^T$, and $\delta_k = [\delta_k^1~~\ldots~~\delta_k^{m_k}]^T$,that track if a node is on the top or bottom layer. Then \eqref{eq:multi-1}-\eqref{eq:multi-2} can be written compactly as
\begin{multline}
    C^p_k \dot{x}_k(t) = -(L_k + H_k) x_k(t) +  \\ h_\infty\sigma_kT_\infty + \kappa_s\delta_kT_s + \tilde B_k(t) u_k(t),
\end{multline}
where $L_k = L(\mc{G}_k)$ is the graph Laplacian (with weights $\kappa^{ij}$) \cite[Section 2.3.3]{mesbahi2010graph}, $H_k = h_\infty \diag(\sigma_k)+\kappa_s \diag(\delta_k)$, $C^p_k$ is a diagonal matrix of heat capacities, and $\tilde B_k(t) = [b^1_k(t)~~\ldots~~b^{m_k}_k]^T$. This can be written as an linear time varying (LTV) system in the standard form
\begin{subequations} \label{eq:full_order_model}
\begin{gather} 
    \dot{x}_k(t) = A_k x_k(t) + B_k(t)u_k(t) + d_k\\
    y_k(t) = B_k(t)^T x_k(t)
\end{gather}
\end{subequations}
where $A_k = -(C^p_k)^{-1}(L_k + H_k)$, $B_k = (C^p_k)^{-1}\tilde B_k$ and $d_k = (C^p_k)^{-1} (h_\infty\sigma_kT_\infty + \kappa_s\delta_kT_s)$. The next layer operator $\mathcal{S}_k$ is then given by the layer-to-layer update
\begin{equation}
    x_{k+1}(0) = \begin{bmatrix}
        e^{A_k \tau_c} \\ 0
    \end{bmatrix} x_k(\tau_k) + \begin{bmatrix}
        A_k^{-1} (e^{A_k \tau_c}  - I)d_k\\
        T_s \mathbf{1}
    \end{bmatrix},
\end{equation}
where $\mathbf{1}$ is a vector of all ones, so that the temperatures evolves according to \eqref{eq:full_order_model} with $u(t) = 0$ during the recoating period $\tau_c$ then new powder with temperature $T_s$ is added.

\subsection{Layer Merging Reduced-order Approximations}
The dimension, and thus the computational complexity, of the full-order model \eqref{eq:full_order_model} grows linearly with the number of layers. This makes it challenging to use \eqref{eq:full_order_model} for control design since SLM parts commonly have anywhere from $10$ to $1000$ or more layers.

\begin{figure}[htbp]
    \centering
    \includegraphics[width=0.95\columnwidth]{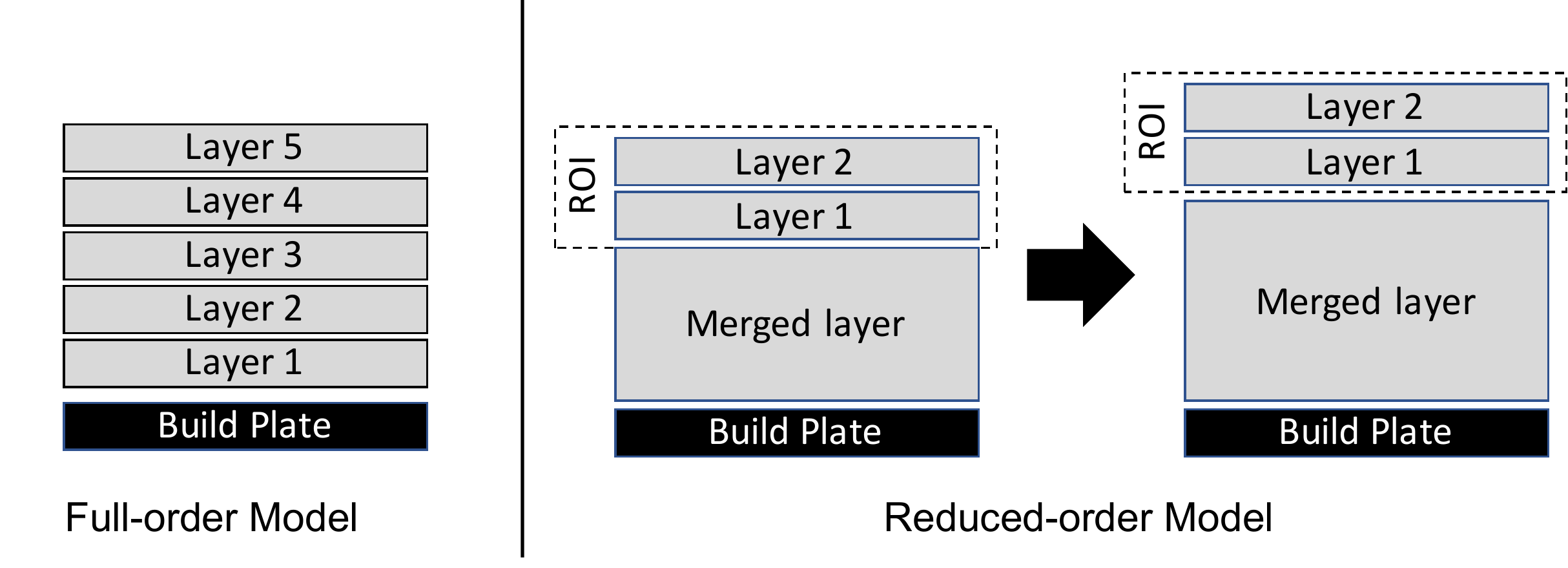}
    \caption{When a new layer is added to the ROM the bottom most layer in the region of interest (ROI) is incorporated into the merged layer before a new power layer is added to the top.}
    \label{fig:roi_schematic}
\end{figure}

To overcome this challenge, we propose a simple model reduction technique. First, we select a layer Region of Interest (ROI) $\gamma$, e.g., $\gamma = 2$ for the top $2$ layers. Then we combine all other layers into a single ``merged layer'', as illustrated in Figure~\ref{fig:roi_schematic}. This leads to a reduced graph $\tilde{\mc{G}}_k = (\tilde{\mc{V}}_k, \tilde{\mc{E}}_k,\tilde{\kappa}_k,\tilde{\ell}_k)$ with $\tilde m_k = |\tilde{\mc{V}}_k|$ nodes; the first $\tilde n_k$ nodes belong to the merged layer. The state of the ROM is 
\begin{equation}
\tilde x_k = [\tilde T^1~~\ldots~~\tilde T^{\tilde n_k}~~\ldots~~\tilde T^{\tilde m_k}]^T.
\end{equation}
The dynamic equations are identical to the full-order case but are based on the reduced graph $\tilde{\mc G}_k$, i.e.,
\begin{subequations} \label{eq:reduced_order_model}
\begin{gather} 
    \dot{\tilde x}_k(t) = \tilde A_k \tilde x_k(t) + \tilde B_k(t)u_k(t) + \tilde d_k\\
    y_k(t) = \tilde B_k(t)^T \tilde x_k(t)
\end{gather}
\end{subequations}
where $\tilde A_k = -(\tilde C^p_k)^{-1}(\tilde L_k + \tilde H_k)$ with weighted graph Laplacian $\tilde L_k = L(\tilde{\mc{G}}_k)$, diagonal matrix of heat capacities $\tilde C^p_k$ and so on.

Adding a layer in the ROM requires incorporating a layer from the ROI into the merged layer. This process is illustrated graphically in Figure~\ref{fig:roi_schematic} and is done so as to ensure energy conservation. Let $\tilde x_k^\ell$ indicate the temperatures of layer $\ell$ in the ROM with $\ell = 0$ denoting the merged layer. Then the layer transition operator $\tilde {\mathcal{S}}_k$ is given by
% \begin{subequations}
\begin{align}
    &\tilde x_k(\tau_k + \tau_c) = e^{\tilde A_k \tau_c} \tilde x_k(\tau_k) + \tilde A_k^{-1} (e^{\tilde A_k \tau_c}  - I)\tilde d_k \nonumber\\
    &\tilde x_{k+1}^0(0) = (\tilde C_k^0 + \tilde C_k^1)^{-1}(\tilde C^0_k \tilde x_k^0(\tau_k+\tau_c) + \tilde C^1_k \tilde x_k^1(\tau_k+\tau_c))\nonumber\\
    &\tilde x_{k+1}^i(0) = \tilde x_k^{i+1}(\tau_k+\tau_c)~~ i = 1,\ldots,\gamma -1\nonumber\\
    &\tilde x_{k+1}^\gamma(0) = \mathbf{1}T_s \label{eq:romTrans}
\end{align}
% \end{subequations}
where $\tilde C_k^i$ is the diagonal matrix of heat capacities for layer $i$. 
The layer above the merged layer is incorporated into the merged layer and the remaining layers are modeled in the same way as the full-order model (FOM).

\begin{figure}[htbp]
    \centering
    \includegraphics[width=0.95\columnwidth]{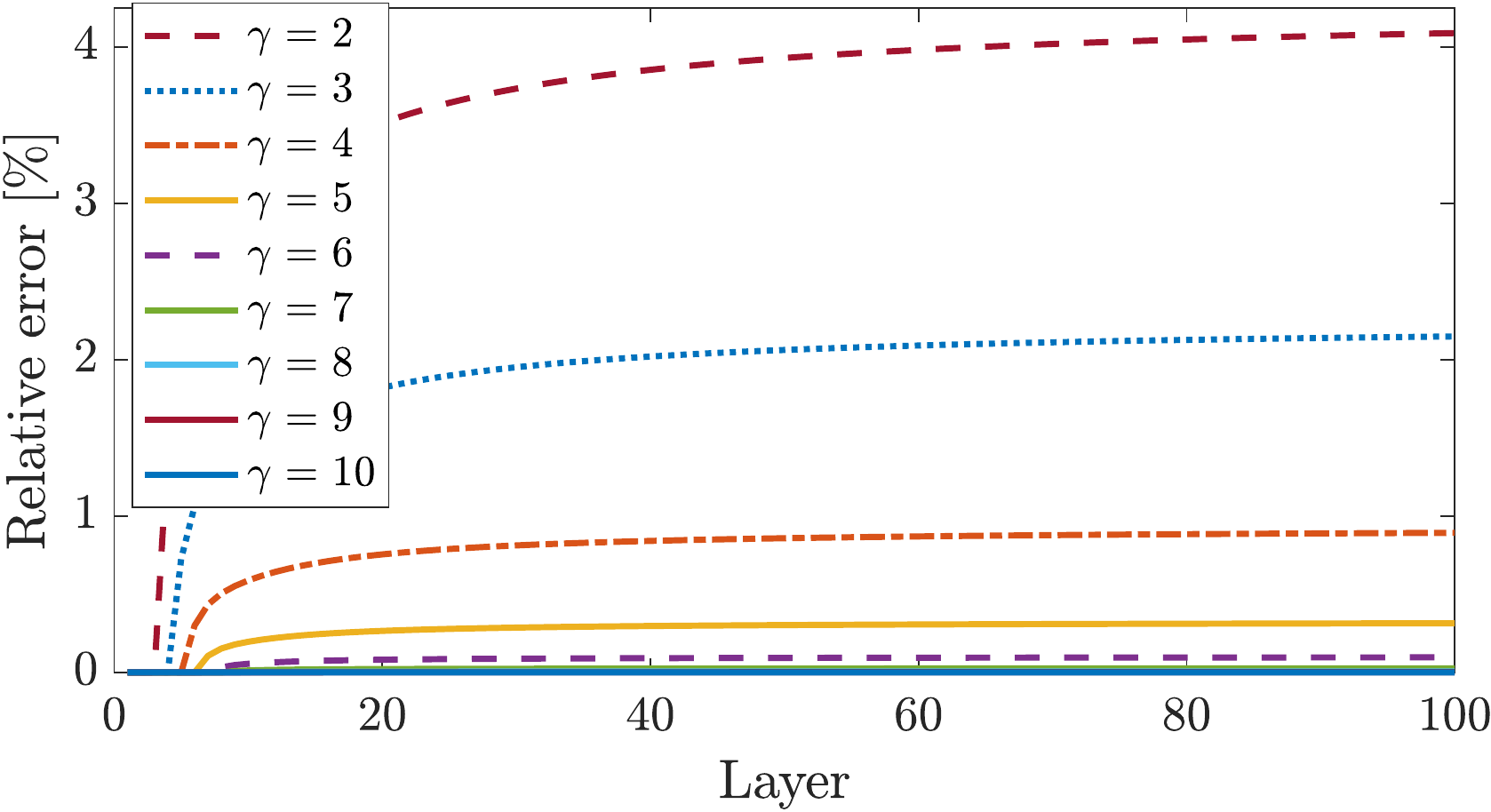}
    \caption{The model-reduction error increases with the total number of layers in the part but eventually stabilizes. If the ROI is greater than the total number of layers, the model is exact and the relative error is zero. The FOM is well approximated by the ROM even with a small ROI.}
    \label{fig:rom_error}
\end{figure}

To quantify the error between the FOM and ROM, we performed a simulation of a straight path using the parameters in Section~\ref{ss:simulations}. The relative error is calculated using the average temperature error (between the FOM and ROM, normalized by the FOM) integrated over the top layer. The ROM approximates the FOM well even with a small ROI as shown in Figure~\ref{fig:rom_error}.

\section{In-layer State Feedback Control} \label{ss:feedback-control}
As illustrated in Figure~\ref{fig:block-diagram}, a typical SLM control architecture consists of two feedback loops; there is usually a significant timescale separation between the two loops (hence the utility of the design). Since the L2L control loop adjusts the control input to be executed in an open-loop for a subsequent layer, it cannot provide in-layer feedback adjustments for improved robustness to disturbances. On the other hand, an in-layer controller operates at high sampling rates (e.g., $1$ to $10$ kHz), which limits the computational complexity of suitable methods. 

In this paper, our objective is to use the laser power input $u_k$ to drive the output $y_k$ to the desired reference trajectory $y^d_k$ via in-layer control. Since the in-layer controller needs to be executable within milli or microseconds, we propose an architecture based on the discrete-time LQR\footnote{We select an LQR architecture instead of a PID controller due to the LQRs well-known robustness properties and to better exploit our model of the SLM process.} and the presented ROM. This leads to an affine feedback law where the gain matrices can be computed offline and only vector addition and matrix multiplication operations are required online. These can be easily parallelized and implemented on high-speed digital hardware, such as a field-programmable gate array (FPGA).

We assume full-state feedback for simplicity. In a typical SLM system, a combination of thermal cameras and pyrometers are available and could be used to measure the temperature distribution of the surface, which in turn may be used to construct an observer for the reduced state. The ROM state is observable from surface temperature measurements\footnote{Output feedback using only temperature measurements in a region around the laser could be useful in practice and is a subject for future work.} and implementing an observer would require only matrix-vector operations, similar to the LQR controller. 

Our proposed controller is based on an exact discretization of the ROM \eqref{eq:reduced_order_model} with sampling period $h > 0$. The resulting dynamics are
\begin{subequations} \label{eq:discrete-rom}
\begin{gather}
\tilde x_k[l+1] = \bar{A}_k \tilde x_k[l] + \bar B_k[l] u_k[l] + \bar d_k\\
y_k[l] = \bar C_k[l] \tilde x_k[l]
\end{gather}
\end{subequations}
where $l$ is the discrete-time index (i.e., $t = l h$), $\bar A_k = e^{\tilde A_k h}$, $\bar B_k[l] = \tilde{A}_k^{-1}(\bar A_k - I)\tilde B_k(lh)$, $\bar d_k = \tilde{A}_k^{-1}(\bar A_k - I) \tilde d_k$ and $\bar C_k[l] = \tilde B_k^T(lh)$.

We define our controller using the solution of the linear-quadratic regulator (LQR) problem
\begin{alignat}{2} \label{eq:LQR_OCP}
\underset{u_k}{\mathrm{min}}& \quad  &&\sum_{l=0}^{N_k} \|y_k[l] - y^d_k[l]\|_Q^2 + \|u_k[l]\|_R^2\\
\mathrm{s.t.}& &&~\tilde x_k[l+1] = \bar{A}_k \tilde x_k[l] + \bar B_k[l] u_k[l] + \bar d_k,~~l=0,...,N_k \nonumber\\
& &&~\tilde x_k[0] = \tilde x_k(0) \nonumber
\end{alignat}
where $N_k = \tau_k/h$ is assumed to be an integer. The optimal control input can be written in state feedback form
\begin{equation} \label{eq:state-feedback}
    u_k[l] = -K_k[l] \tilde x_k[l] + K^v_k[l] v[l+1] - K^d_k \bar d_k,
\end{equation}
and consists of feedback, feedforward, and disturbance rejection terms \cite{singh2017extended}. The feedback gain $K_k$ is given by 
\begin{equation*}
    K_k[l] = (\bar B_k[l]^T P_k[l+1] \bar B_k[l] + R)^{-1} \bar B_k[l] P_k[l+1] \bar A_k[l]
\end{equation*}
where $P_k$ is the solution of the Riccati recursion
\begin{equation} \label{eq:riccati}
    P_k[l] = \bar A_k^T[l] P_k[l+1](\bar A_k[l] - \bar B_k[l] K_k[l]) + \bar C_k[l] Q \bar C_k^T[l],
\end{equation}
for $l = 1,\ldots, N_{k}-1$ with boundary condition $P_k[N_k] = \bar C_k[N_k] Q \bar C_k[N_k]^T$. The feedforward term consists of a gain
\begin{equation*}
    K_k^v[l] = (R + \bar B_k^T[l] P_k[l+1] \bar B_k[l])^{-1} B_k^T[l]
\end{equation*}
and an auxiliary sequence
\begin{equation} \label{eq:aux_rec}
    v_k[l] = (\bar A_k[l] - \bar B_k[l] K_k[l])^T v_k[l+1] + \bar C_k^T[l] Q y^d_k[l]
\end{equation}
for $l = 1,\ldots, N_{k}-1$ with terminal condition $v_k[N_k] = \bar C_k^T[N_k] Q y^d_k[N_k]$. The disturbance compensation term is 
\begin{equation*}
    K_k^d[l] = K_k[l](P_k[l] - \bar C_k[l]Q \bar C_k^T[l])^{-1}V_k[l]
\end{equation*}
where $V_k$ is defined by the backwards recursion
\begin{equation} \label{eq:V_rec}
    V_k[l] = (\bar A_k[l] - \bar B_k[l] K_k[l])^T(P_k[l+1] + V_k[l+1])
\end{equation}
for $l = 1,\ldots, N_{k}-1$ with $V_k[N_k] = 0$.

In practice, the laser power is limited to a range $\mc{U} = [0,p_{\max}]$ for some maximum power $p_{\max}$. We accommodate this constraint by setting
\begin{equation} \label{eq:feedback-law}
    u_k[l] = \Pi_\mc{U}(-K_k[l] \tilde x_k[l] + K^v_k[l] v[l+1] - K^d_k \bar d_k),
\end{equation}
where $\Pi_\mc{U}$ denotes Euclidean projection onto the set $\mc{U}$. Ideally, the input constraint $u(t)\in \mc{U}$ would be incorporated directly into the optimization problem \eqref{eq:LQR_OCP}, which would then be solved repeatedly, resulting in an model predictive control (MPC) like structure, e.g.,~\cite{zuliani2021batch}. However, the computational cost of the resulting controller is likely prohibitive for high loop-rates. Moreover, the dimensionality of the system in practical applications is too large for explicit MPC \cite{bemporad2002explicit} even after reduction. We therefore opt for a suboptimal projection approach.

The overall process is summarized in Algorithm~\ref{alg:closedloop} and consists off an offline phase where the feedback gains and feedforward terms are computed for each layer and an online phase where they are combined with measurements.
\begin{algorithm}
    \caption{In-layer Feedback}
    \label{alg:closedloop}
    \begin{algorithmic}[1]
    \State Select $y_d^k, Q, R, p_{\max}$
        \For{$k = 1,\ldots, N$} \Comment{Offline phase}
            \State Construct $\bar A_k, \bar B_k, \bar C_k, \bar d_k$ using \eqref{eq:reduced_order_model} and \eqref{eq:discrete-rom}
            \State Compute $P_k, v_k$, and $V_k$ using \eqref{eq:riccati}, \eqref{eq:aux_rec}, and \eqref{eq:V_rec};
            \EndFor
            \For{$k = 1,\ldots, N $} \Comment{Online phase}
                \For{$l = 0, \ldots, N_k$} 
                \State Measure/estimate $\tilde x_k[l]$ 
                \State Compute $u_k[l]$ with \eqref{eq:feedback-law}, apply to the system \eqref{eq:full_order_model}
                \EndFor
            \EndFor
    \end{algorithmic}
\end{algorithm}
\vspace{-0.25cm}

\section{Simulation case study} \label{ss:simulations}

We consider a part with $N = 20$ layers and take our build volume to be an $L_x \times L_y$ box with $n_x = n_y$ cells per axis. The layer thickness is $\delta_z$, the ambient and substrate temperatures are $T_\infty$ and $T_s$, and the convection coefficient is $h_\infty$. The beam has radius $R$ and max power $p_{\max}$. 
We use Stainless Steel 316L, which is commonly used in SLM processes, as our printing material. It has a specific heat capacity of $c_p$, convection coefficients $k_p$ and $k_d$ in powder and solid form respectively, porosity $\epsilon$ which is used to calculate the heat capacity of the powder as in \cite{roy2018heat}, and absorption coefficient $\alpha$. The numerical values for the process parameters are given in Table~\ref{tab:params}. 

\begin{figure}[htbp]
    \centering
    \includegraphics[width=0.95\columnwidth]{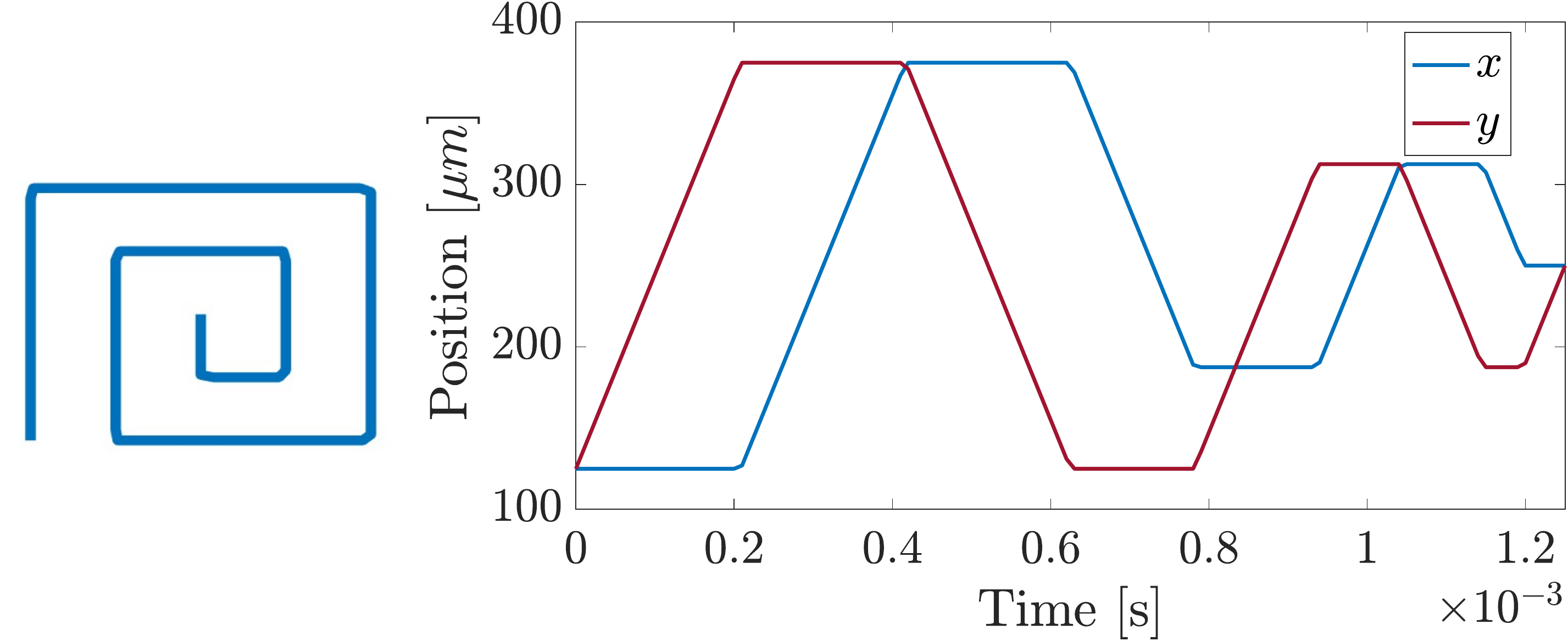}
    \caption{The square spiral laser path (left) and trajectories (right). The powder bed is $500 \mu$m$\times 500 \mu$m.}
    \label{fig:paths}
    \vspace{-0.15cm}
\end{figure}

We consider the square spiral trajectory illustrated in Figure~\ref{fig:paths}, scanned at $1.2~m/s$. As an output reference, we use a constant value of $y^d_k = 1700~K$ which is just above the melting temperature of SS 316L. We use the weights $Q = R = 1$. Regulating the average temperature under the laser helps stabilize the meltpool, leading to more consistent prints~\cite{shkoruta2021real}. 

\begin{table}[hb]
\centering
\vspace{-0.25cm}
\caption{Model Parameters}
\label{tab:params}
\begin{tabular}{|c c|c c|} \hline
Parameter & Value & Parameter & Value \\ \hline \hline
$L_x,L_y$ & $500~[\mu m]$ & $n_x,n_y$ & $25$ \\ \hline
$\delta_z$ & $50~[\mu m]$ & $T_\infty$ & 300 [K]\\ \hline
$T_s$ & $900$ [K] & $h_\infty$ & $10~\left[\frac{W}{m^2K}\right]$\\ \hline
$p_{\max}$ & $50$ [W] & $R$ & $60~[\mu m]$\\ \hline
$\tau_k$ & $1.25$ [ms] & $\tau_c$ & $1.25$ [ms]\\ \hline
$h$ & $10~[\mu s]$ & $c_p$ & $4.25 \times 10^6~\left[\frac{J}{m^3K}\right]$\\ \hline
$k_p$ & $0.5~\left[\frac{W}{mK}\right]$ & $k_d$ & $20~\left[\frac{W}{m K}\right]$\\ \hline
$\epsilon$ & $0.5$ & $\alpha$ & $0.42$\\ \hline
\end{tabular}
\end{table}

\begin{figure}[htbp]
    \centering
    \includegraphics[width=0.95\columnwidth]{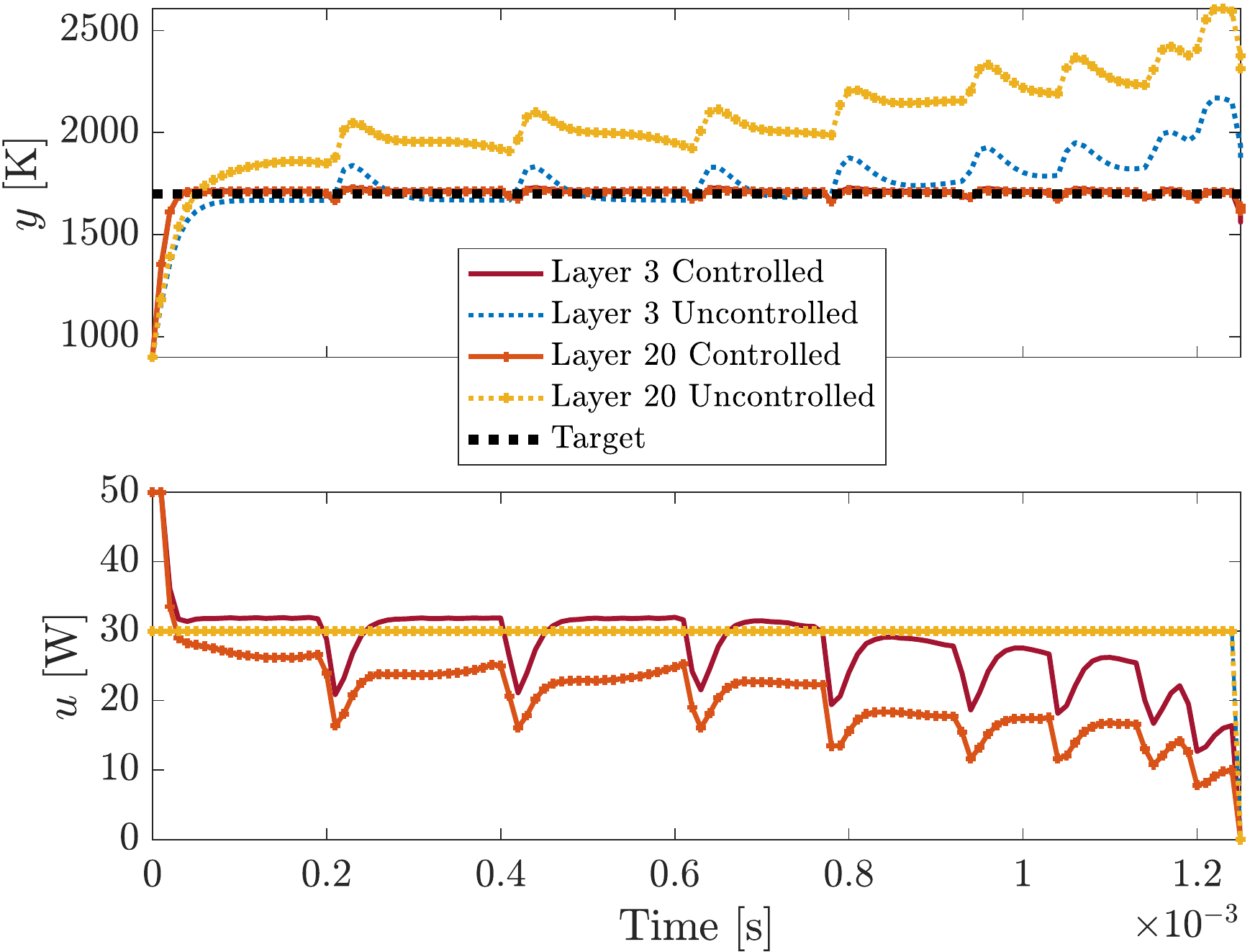}
    \caption{When uncontrolled, significant amounts of energy accumulate in the system as the number of layers increases and may cause thermal defects. The controller reduces the laser power as the build progresses to prevent this buildup.}
    \label{fig:vs_uncontrolled}
    \vspace{-0.25cm}
\end{figure}

Figure~\ref{fig:vs_uncontrolled} compares the controlled and uncontrolled responses of the system for several different layers. The controller is able to successfully adjust input power to keep the average temperature near the laser around its setpoint. This clearly illustrates the potential for closed-loop control to regulate the temperature of the part and prevent gradual heat buildup. Heat buildup is evident in the uncontrolled trajectories and may cause cracking, thermal stress, or other defects in the final part.

\begin{figure}[htbp]
    \centering
    \vspace{-0.25cm}
    \includegraphics[width=0.95\columnwidth]{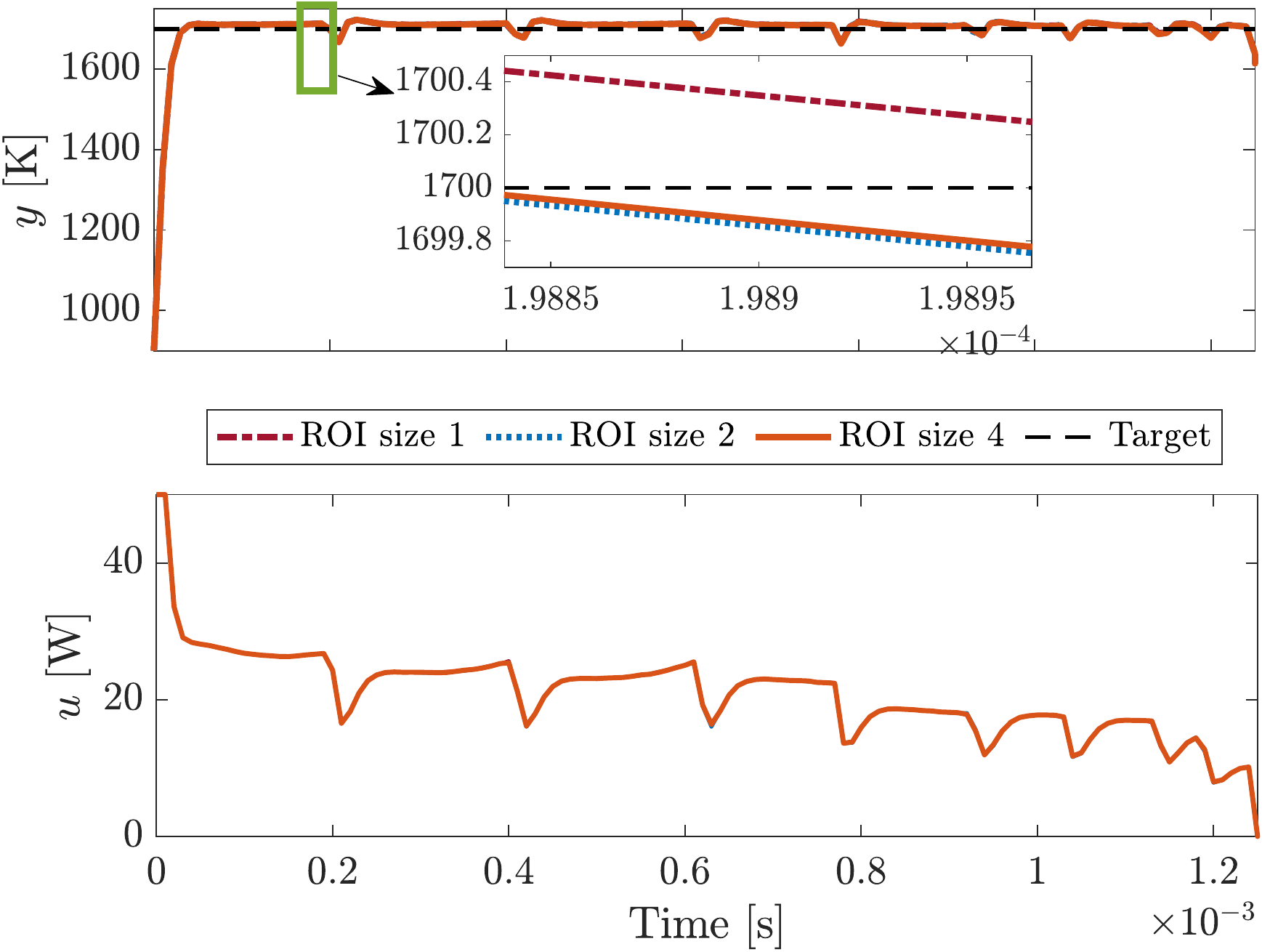}
    \caption{When placed in closed-loop, the depth of the region of interest (ROI) has a negligible impact on the closed-loop performance of the controller. Trajectories from layer $19$.}
    \label{fig:roi_comp}
    \vspace{-0.25cm}
\end{figure}

Figure~\ref{fig:roi_comp} illustrates the impact of using different regions of interest depths on the closed-loop response. The benefit gained from increasing the ROI from $\gamma = 1$ to $\gamma = 4$ is negligible, demonstrating that the considering one previous layer is sufficient for control. This is beneficial since it reduces computational complexity; moreover, as it is possible to measure the temperature field of the top surface, observer design in the case of output feedback is greatly simplified. 

Overall, these simulations illustrate the potential impact even a simple controller could have on the reliability and consistency of multi-layer SLM processes and that the input-output relationship between the temperature near the laser and the laser power can be captured well enough for control design using relatively low-order models (e.g., one or a few top layers in the ROI with the merged layer dynamics for the rest of the part).
Incorporating additional physical phenomena (e.g., melt pool dynamics, phase changes of the material) to improve the fidelity of the model may affect the closed-loop system performance with reduced ROI. Computational studies to quantify the sensitivity to ROI with higher fidelity models are subjects for future work. 
\begin{rmk}
Figure~\ref{fig:roi_comp} demonstrates the robustness of our proposed control strategy to model mismatch caused by model order reduction. We expect a similar degree of robustness to other sources of mismatch such as uncertain parameters.
\end{rmk}

\section{Conclusion} \label{ss:conclusion}
This paper extends a graph-based finite element approach for modeling a selective laser melting process to consider multiple layers and proposes a simple but effective model order reduction technique to arrive at a practical control-oriented thermal model of a multi-layer process. Based on this model, we propose an LQR-based controller that regulates an average temperature by varying the laser power and is simple enough to allow the construction of high-frequency feedback loops. Finally, we illustrate the potential of the controller for mitigating layer-to-layer heat buildup which may cause cracking, thermal stress, or other defects in finished parts. Future directions include: integrating our multi-layer thermal model with a more sophisticated model of the melt pool, adding an observer for the reduced-order state, validating our approach on experimental hardware and/or high-fidelity simulations, augmenting our models with data-driven terms, and optimally incorporating constraints into our controller formulation.

\bibliography{slm}
\end{document}